\documentclass[portrait,11pt]{article}
\usepackage[sectionbib,round]{natbib}
\usepackage{times,amsmath,amsfonts}
\usepackage[dvips]{graphicx,graphics}

\usepackage{amsfonts,amsmath,amssymb}
\usepackage{url}
\usepackage{hyperref}
\hypersetup{colorlinks,%
citecolor=red,%
filecolor=black,%
linkcolor=red,%
urlcolor=red,%
pdftex}

\newtheorem{theorem}{Theorem}
\newtheorem{definition}{Definition}

\newcommand{\bqn}{\begin{eqnarray*}}
\newcommand{\eqn}{\end{eqnarray*}}
\newcommand{\bq}{\begin{eqnarray}}
\newcommand{\eq}{\end{eqnarray}}

\begin{document}
%\pagenumbering{arabic}
\pagenumbering{gobble}

\title{Graph Theory in Brain Networks}
\author{Moo K. Chung \\
University of Wisconsin-Madison\\
\quad \tt{mkchung@wisc.edu}}
\maketitle

Recent developments in graph theoretic analysis of complex networks have led to deeper understanding of brain networks. Many complex networks show similar macroscopic behaviors despite differences in the microscopic details \citep{bullmore.2009}. Probably two most often observed characteristics of complex networks are scale-free and small-world properties \citep{song.2005}. In this paper, we will explore whether  brain networks follow scale-free and small-worldness among other graph theory properties. This short paper duplicates the chapter 3 of book \citet{chung.2019.BNA} with additional errors corrected with a new figure.

\section{Trees and graphs}
\index{trees}
\index{graph!trees}

Many objects and data can be represented as networks. Unfortunately, networks can be very complex when the number of nodes increases. Trees, which can be viewed as the backbone of the networks, are often used as a simpler representation of the networks.

\begin{definition}
An (undirected) {\em graph} is an ordered pair $G = (V, E)$ with a node (vertex) set  $V$ and  edge (link) set  $E$ connecting the nodes. The vertices belonging to an edge are called the ends or end vertices of the edge. A vertex may exist in a graph and not belong to an edge. The size of the graph is usually determined by the number of vertices  $|V|$ and  the number of edges $|E|$. 
\end{definition}

A (weighted) graph with $p$ nodes is often represented as a connectivity matrix $C=(c_{ij})$ of size $p \times p$, where $c_{ij}$ are usually referred to as edge weights. A {\em binary graph} is a graph with binary edge weights, i.e., $(0,1)$. An undirected graph yields a symmetric connectivity matrix. For instance, given normalized and scaled data matrix $X,$ $Y$ of size $n \times p$, $corr(X,X)$ gives an undirected weighted graph while $corr(X,Y)$ gives a directed weighted graph.

\begin{definition} 
A {\em tree} is an undirected graph, in which any two nodes are connected by exactly one path. 
Nodes with only one edge in a tree are referred to as {\em leaves} or {\em leaf nodes}  \citep{stam.2014}. The number of leaves is the {\em leaf number}. A {\em binary tree} is a tree, in which each node has at most two connecting nodes called left or right children. A {\em forest} is a disjoint union of trees. 
\end{definition}

In brain imaging, we often deal with brain cortical mesh vertices. To distinguish such vertices from the vertices of graphs, we will simply use the term {\em nodes}. Instead of edges, the term {\em links} is also used. The term {\em tree} was coined in 1857 by the British mathematician Arthur Cayley. 
Node degrees are possibly the most often used feature in characterizing the topology of a tree. In a tree, there is only one path connecting any two nodes. Thus, the average {\em path length} between all node pairs is easy to compute. 

\begin{theorem}
For a tree, we have $|V| - |E| = 1$. For a forest, the total number of trees is give by $|V| - |E|$.
 \end{theorem}
The proof is straightforward enumeration of edges with respect to the nodes.

\begin{definition}
A {\em rooted tree} is a tree, in which one node is designated the root. In a rooted tree, the {\em parent} of a node is the node connected to it on the path to the root. Given node $x$, a {\em child} of $x$ is a node of which $x$ is the parent. A {\em sibling} to node $x$ is any other node on the tree which has the same parent as $x$. 
\end{definition}

\begin{definition}
In a rooted tree, the {\em ancestors} of a node are the nodes in the path from the node to the root, excluding the node itself and including the root. The {\em descendants} of node $x$ are those nodes that have $x$ as an ancestor.
\end{definition}

\begin{definition}
The {\em height} of a node in a rooted tree is the length of the longest downward path from a leaf from that vertex. The {\em depth} of a node is the length of the path to its root. 
\end{definition}

We can show that every node except the root has a unique parent. The root has depth zero, leaves have height zero.

\section{Minimum spanning trees}
\index{trees!minimum spanning trees}
\index{minimum spanning trees}

The trees are mainly used as a simpler representation of more complex graphs. The computation of graph theory features for trees is substantially easier that that of graphs. Among all trees, minimum spanning trees (MST) are most often used in practice. 

\begin{definition}
A {\em spanning tree} of a graph $G$ is a tree whose node set is identical to the node set of $G$. 
The {\em minimum spanning tree} (MST) is a spanning tree whose sum of edge weights is the smallest. 
\end{definition}

\begin{theorem}If all the edge weights are unique in a graph, there exist one unique MST.
\end{theorem}
{\em Proof.} We prove by contradiction. Suppose there are two MSTs $M_1=(V, E_1)$ and $M_2=(V,E_2)$. Since $E_1 \neq E_2$,  there are edges that belongs to one but not both. Among such edges, let $e_1$ be the one with the least weight. Without loss of generality, assume $e_1 \in E_1$. Adding $e_1$ to $M_2$ will create a cycle $C$. Since $M_1$ has no cycle, $C$ must have an edge $e_2$ not in $M_1$, i.e., $e_2 \in E_2$. Since $e_1$ is the least edge weight among all the edges that  belong to one but not both, the edge weight of $e_1$ is strictly smaller than the weight of $e_2$. Replacing $e_2$ with $e_1$ will yield a new spanning tree with wights less than that of $M_2$, which is a contradiction. Thus, there must be only one unique MST. $\square$

\begin{figure}
\centering
\includegraphics[width=0.9\linewidth]{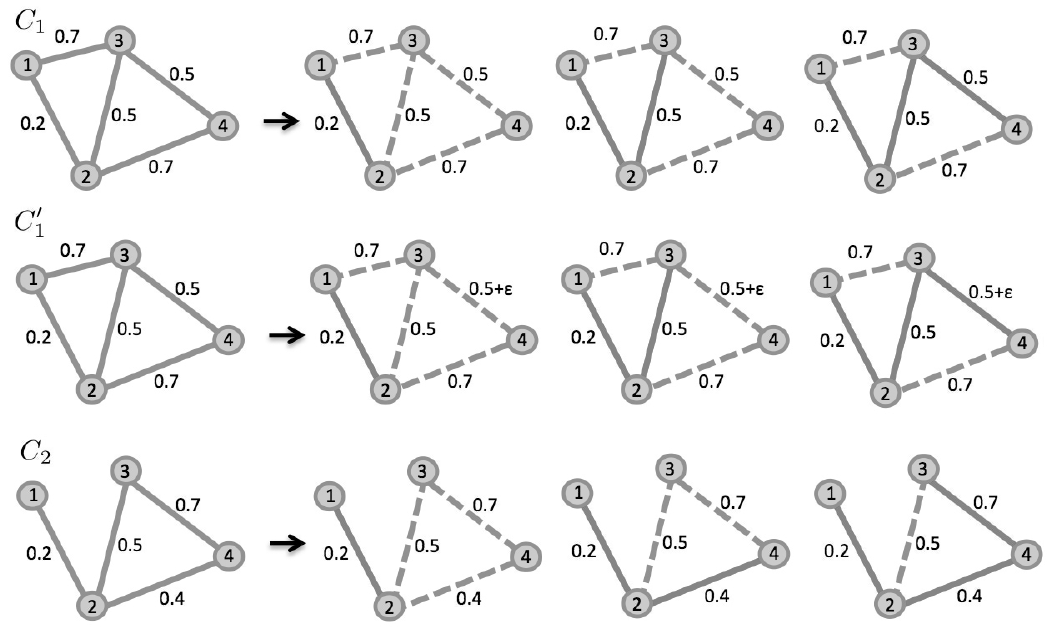}
\caption{Minimum spanning tree (MST) construction using Kruskal's algorithm. $C_1$: The edge weights of MST are 0.2, 0.5, 0.5. Even though we have identical edge wights 0.5, we can numerically still have a unique MST by putting infinitesimally small weight $\epsilon$, which results in $C'_1$. $C_2$: The edge weights of MST are 0.2, 0.4, 0.7.}
\label{fig:NI-MSTschematic}
\end{figure}

MST is often constructed using Kruskal's algorithm \citep{lee.2012.TMI}. Kruskal's algorithm is a greedy algorithm with runtime $O (|E| \log |E|) = O (|E| \log |V|)$. Note $|E|$ is bounded by $|V|(|V|-1)/2$. The  run time is equivalent to the run time of sorting the edge weights which is also $O (|E| \log |E|)$. The algorithm starts with an edge with the smallest weight. Then add an edge with the next smallest  weight. This sequential process continues while avoiding a loop and  generates a spanning tree with the smallest total edge weights (Figure \ref{fig:NI-MSTschematic}). Thus, the edge weights in MST correspond to the oder, in which the edges are added in the construction of MST. It is known that the single linkage hierarchical clustering is related to Kruskal's algorithm of MST \citep{gower.1969}.

%\begin{example}
%\label{eq:perm-C1C2}
Consider two graphs $C_1$ and $C_2$ shown in Figure \ref{fig:NI-MSTschematic}. In Matlab, the edge weights can be encoded as weighted adjacency matrices
\begin{verbatim}
C1= [0 0.2 0.7 0
     0.2 0 0.5 0.7
     0.7 0.5 0 0.5
     0 0.7 0.5 0]

C2= [0 0.2 0 0
     0.2 0 0.5 0.4
     0.7 0.5 0 0.7
     0 0 0.4 0]
\end{verbatim}
The diagonals are assigned value zero. The minimum spanning tree is obtained using {\tt graphminspantree.m} which inputs a sparse connectivity matrix.
\begin{verbatim}
C1=sparse(C1)
C2=sparse(C2)
[treeC1,pred]=graphminspantree(C1)
[treeC2,pred]=graphminspantree(C2)

treeC1 =
   (2,1)       0.2000
   (3,2)       0.5000
   (4,3)       0.5000
 
 treeC2 =
   (2,1)       0.2000
   (3,2)       0.5000
   (4,3)       0.4000  
\end{verbatim}
By connecting all the nonzero entries in the output {\tt treeC1} and {\tt treeC2}, which are given as sparse matrices, we obtain MST.

When some edges have equal wight as shown in Figure \ref{fig:NI-MSTschematic}, we may not have a unique MST. Thus, there are at most $q!$ possible MSTs. Given a binary graph with $p$ nodes, any spanning tree is a MST. Given a weighted graph with $q$ identical edge weights, we can assign infinitesimally small weights $\epsilon,$ $2\epsilon,$ $\cdots, q\epsilon$ to the identical edge weights. This results in a graph with unique edge weights and there exists a single MST. There are at most $q!$ ways to assign infinitesimally weights to $q$ edges. For a complete graph with $p$ nodes, the total number of spanning tree  is $p^{p-2}$. For an arbitrary binary graph, the total number can be calculated in polynomial time as the determinant of a matrix from Kirchhoff's theorem \citep{chaiken.1978}. 

\begin{figure}
\centering
\includegraphics[width=1\linewidth]{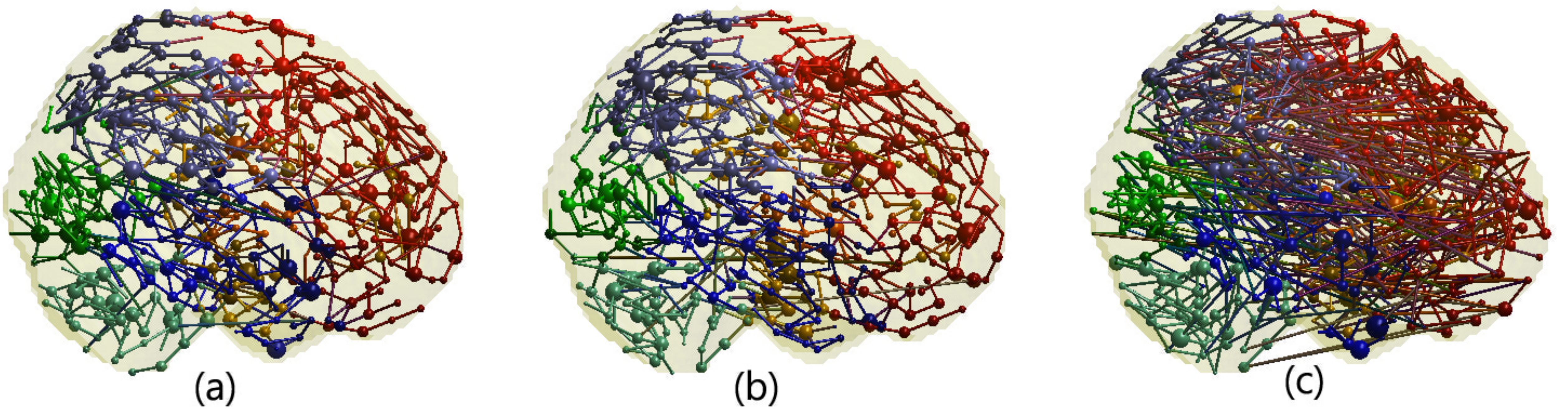}
\caption{MST of the brain networks of (a) attention-deficit hyperactivity disorder, (b) autism spectrum disorder and (c) pediatric control  subects. The number of nodes are 1056 in all networks.  The color represents the 97 regions of interest based on the predefined anatomical parcellations.
Figure was generated by Hyekyoung Lee of Seoul National University \citep{lee.2012.TMI}.}
\label{fig:permute-MST}
\end{figure}

\begin{theorem}
\label{thm:rho'}
Suppose $\rho_1, \cdots, \rho_{p-1}$ are the ordered edge weights of the MST of a graph $G$. 
If $\rho'_1, \cdots, \rho'_{p-1}$ are the ordered edge weights of any other spanning tree of $G$. Then we have
$$\rho_j \leq \rho'_j$$
for all $j$.
\end{theorem}
Theorem \ref{thm:rho'} shows the ordered edge eights of MST is extremely stable relative to any ordered edge weights of a spanning tree and thus, they can be used as stable multivariate feature for quantifying graphs.

\section{Node degree}
\index{graph!degree}
\index{node degree}
\index{degree}

Probably the most important network complexity measure is node degree. Many other graph theory measures are related to node degree \citep{bullmore.2009}. The {\em degree} $k$ of a node is the number of edges connected to it. It measures the local complexity of network at the node (Figure \ref{fig:graph-degreer}). Once we have the adjacency matrix of a network, the node degree can be computed easily by summing up the corresponding rows or columns in the adjacency matrix. The average node degree is simply given in terms of $|E|$ and $|V|$:

\begin{figure}[th!]
\centering
\includegraphics[width=0.8\linewidth]{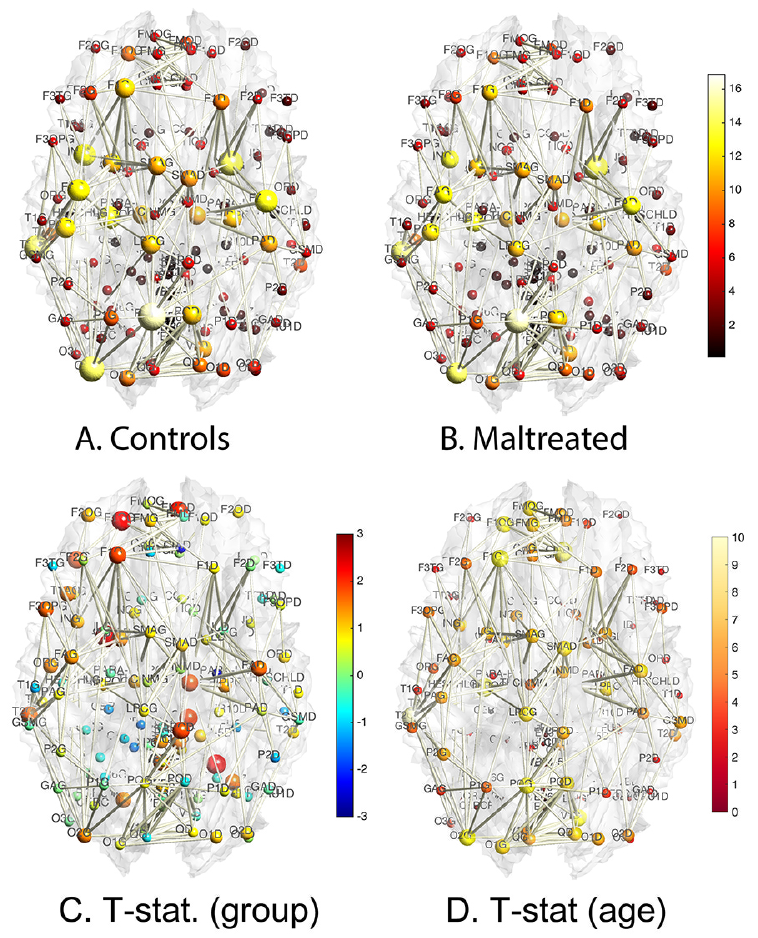}
 \caption{The node size and color correspond to the mean degree of normal controls (A) and maltreated children (B) \citep{chung.2017.BC}.  The edges are the average of the mean degrees of the two nodes threshoded at 0.5. C. The two-sample $t$-statistic of the degree differences (controls - maltreated).   D. $t$-statistic of age effect while accounting for sex and group variables.}
 \label{fig:graph-degreer}
\end{figure}

\begin{theorem} For undirected network, the average degree $\mathbb{E} k$ is 
$$\mathbb{E} k = \frac{2 |E|}{|V|},$$
where $|V|$ is the total number of nodes and $|E|$ is the total number of edges in the graph. 
\end{theorem}
{\em Proof.} The total number of node degree is $2|E|$. Thus, the average node degree is $2|E|/|V|$. $\square$\\

\begin{figure}[t]
\begin{center}
\includegraphics[width=1\linewidth]{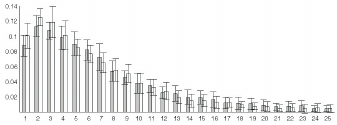}
 \caption{Degree distribution for autistic (white) and control (gray) subjects. Autistic network is more stratified showing higher concentration of low degree nodes (degree 1 to 3) compared to the control subjects (pvalues = 0.024, 0.015, 0.080 respectively) \citep{chung.2010.hbm.epsilon}.}
 \label{fig:connectivity-histogram}
 \end{center}
\end{figure}

\subsection{Scale-free networks} 
\index{scale-free}
\index{graph!scale-free}

 The {\em degree distribution} $P(k)$, probability distribution  of the number of edges $k$ in each node, does not have heavy tails. Figure \ref{fig:connectivity-histogram} shows a study showing degree distribution difference between autistic and normal controls \citep{chung.2010.hbm.epsilon}. The usual two-sample $t$-tests will not work in the tail regions since the variance is too large due to small sample size. Inference on tail regions requires extreme value theory, which deals with modeling extreme events and has seen applications in environmental studies \citep{smith.1989} and insurance \citep{embrechts.1999}. One main tool in extreme value theory is the use of  generalized Pareto distribution in approximating the tail distributions at high thresholds. A standard technique is  to estimate tail regions with parametric models and perform inferences on the parameters of the model fit. For low degrees, since the sample size is usually large, two-sample $t$-test is sufficient.\\

\begin{figure}[t!]
\centering
\includegraphics[width=1\linewidth]{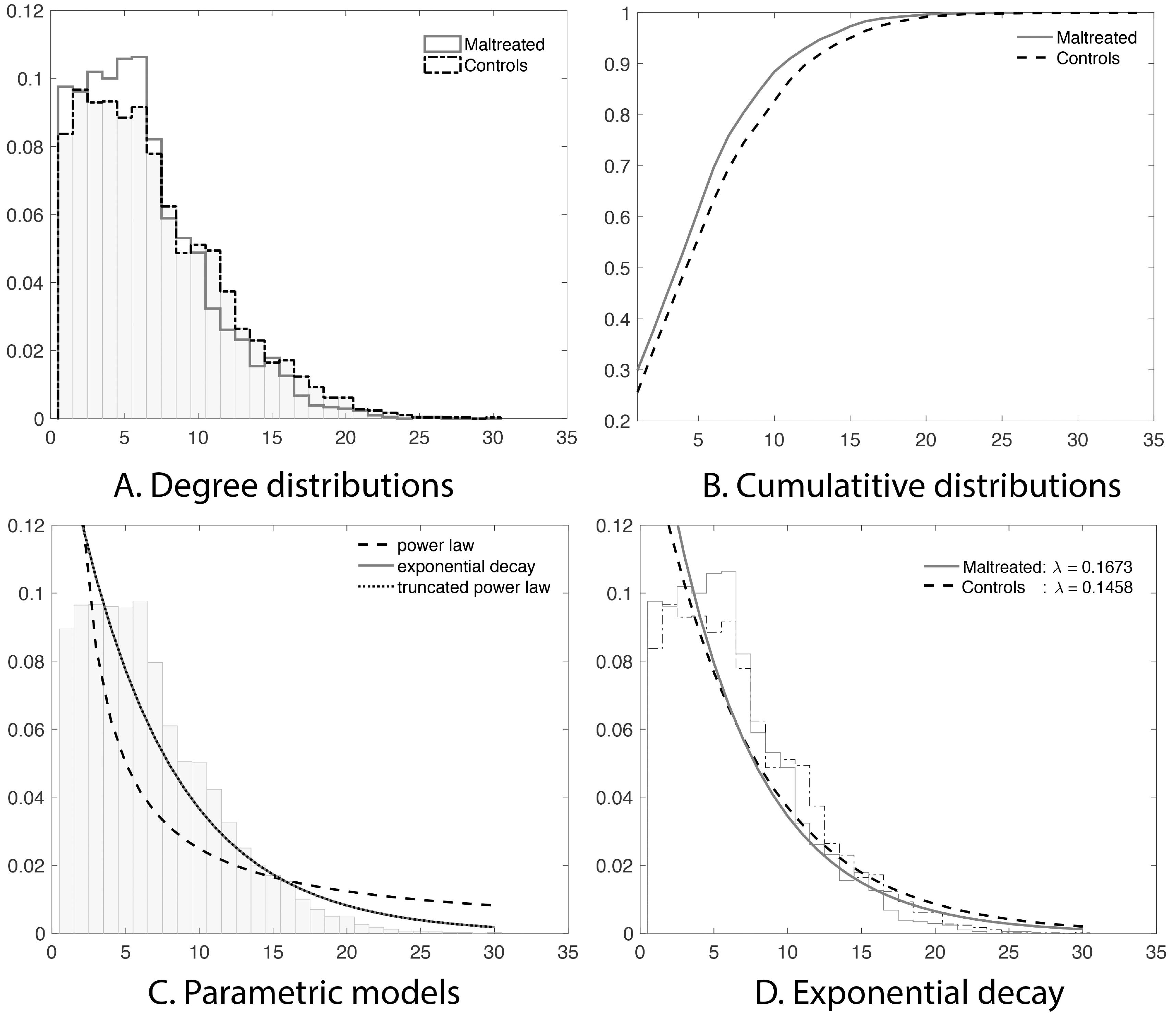}
 \caption{A. Degree distributions of all the subjects combined in each group \citep{chung.2017.BC}. B. The cumulative distribution functions (CDF) of all the subjects in each group. C. Three parametric model fit on the CDF of the combined 54 subjects. %The exponential decay model fit the CDF best and there is slight shape difference between the exponential decay model and the truncated power law. 
 D. The exponential decay model is fitted in each group. The estimated parameters are significantly different ($p$-value $<$ 0.02).}\label{fig:degree}
\end{figure}

The degree distribution $P(k)$ can be represented by a power-law  with a degree exponent $\gamma$ usually in the range $2  <  \gamma < 3$ for diverse networks \citep{bullmore.2009, song.2005}:
$$ P_{p}(k) \sim  k^{-\gamma}.$$
Such networks exhibits gradual decay of tail regions (heavy tail) and are said to be {\em scale-free}. In a scale-free network, a few hub nodes hold together many nodes while in a random network, there are no highly connected hub nodes. The smaller the value of $\gamma$, the more important the contribution of the hubs in the network. 

Previous studies have shown that the human brain network is not scale-free \citep{gong.2009,hagmann.2008, zalesky.2010}. \citet{hagmann.2008} reported that degree decayed exponentially, i.e., 
$$P_{e}(k) \sim e^{-\lambda k},$$
where $\lambda$ is the rate of decay \citep{fornito.2016}. The smaller the value of $\lambda$, the more important the contribution of the hubs in the network. 

\citet{gong.2009} and \citet{zalesky.2010} found the degree decayed in a heavy-tailed manner following an exponentially truncated power law
$$P_{etp}(k) \sim k^{-\gamma}e^{-\lambda k},$$
where $1/\lambda$ is the cut-off degree at which the power-law transitions to an exponential decay \citep{fornito.2016}. This is a more complicated model than the previous two models.

The estimated best model fit can be further used to compare the model fits among the three models. However, existing literature on graph theory features mainly deal with the issue of determining if the brain network follows one of the above laws \citep{fornito.2016,gong.2009,hagmann.2008, zalesky.2010}. However, such model fit was not often used for actual group-level statistical analysis.

\subsection{Estimating degree distributions} 
\index{distributions!degree}
\index{degree distributions}

Directly estimating the parameters from the empirical distribution is challenging due to small sample size in the tail region. This is probably one of the reasons we have conflicting studies. To avoid the issue of sparse sampling in the tail region, the parameters are often estimated from cumulative distribution functions (CDF) that accumulate the probability from low to high-degrees and reduce the effect of noise in the tail region. To increase the sample size further in the tail region, we combine all the degrees across the subjects. For the exponentially truncated power law, for instance, the two parameters $\gamma, \lambda$ are then estimated by minimizing the sum of squared errors (SSE) using the $L_2$-norm between theoretical CDF $F_{etp}$ and empirical CDF $\widehat F_{etp}$:
$$  (\widehat \gamma ,\widehat \lambda)   = \arg \min_{\gamma \geq 0, \lambda \geq 0} \int_{0}^{\infty} \big| F_{etp}(k) - \widehat{F}_{etp}(k) \big|_2^2 \; dk.$$

We propose a two-step procedure for fitting node degree distribution.
The underlying assumption of the two-step procedure is that each subject follows the same degree distribution law but with different parameters. In the first step,  we need to determine which law the degree distribution follows at the group-level. This is done by pooling every subject to increase the robustness of the fit. In the second step, we determine subject-specific parameters.

{\em Step 1.} This is a group-level model fit. Figure \ref{fig:degree}-A  shows the degree distributions of the combined subjects in a group \citep{chung.2017.BC}. To determine if the degree distribution follows one of the three laws, we combine all the degrees across subjects in the group. Since high-degree hub nodes are very rare, combining the node degrees across all subjects increases the robustness of the fit. This results in much more robust estimation of degree distribution. At the group-level model fit, this is possible. 

{\em Step 2.} This is the subject-level model fit. Once we determine that the group follows a specific power law, we fit  the the same model for each subject separately. 

\begin{table}[th!]
\caption{\label{table:hubs}13 most connected hub regions obtained from DTI study comparing normal controls and maltreated children \citep{chung.2017.BC}. AAL regions are sorted in the descending order of the node degree. The controls have more connections without an exception compared to maltreated children.} 
\centering 
\small
\begin{tabular}{|c|c|c|c|c|c|} 
\hline
Label   &    Parcellation Name & Combined & Controls & Maltreated\\
\hline
PQG	&	Precuneus-L & 16.11	&	16.87	&	15.09\\
NLD&	Putamen-R & 14.96	&	15.26	&	14.57\\
O2G	&	Occipital-Mid-L	&	14.44 & 15.52	&	13.00\\
T2G	&	Temporal-Mid-L &14.30	&	15.16	&	13.13\\
HIPPOG	&Hippocampus-L &	13.15	&	13.94	&	12.09\\
FAD	&	Precentral-R & 12.85	&	14.00	&	11.30\\
ING	&	Insula-L & 12.56	&	13.61	&	11.13\\
FAG	&	Precentral-L &12.43	&	13.45	&	11.04\\
PQD	&	Precuneus-R & 12.00	&	12.03	&	11.96\\
PAG	&	Postcentral-L & 11.89	&	12.52	&	11.04\\
NLG	&	Putamen-L & 11.39	&	11.68	&	11.00\\
F1G	&	Frontal-Sup-L & 11.22	&	12.13	&	10.00\\
HIPPOD & Hippocampus-R &	11.15	&	11.90	&	10.13\\
\hline
\end{tabular}  
\end{table}
\normalsize

\subsection{Hub nodes} 
\index{nodes!hub}
\index{hub nodes}

Hubs or hub nodes are defined as nodes with high-degree of connections \citep{fornito.2016}. Table \ref{table:hubs} shows the list of the 13 most connected nodes in a DTI study of maltreated children vs normal controls. The numbers are the average node degrees in each group. All 13 nodes showed higher degree values in the controls without an exception. The probability of this event happening by random chance alone is  $2^{-13} = 0.00012$. This is an unlikely event and we conclude that the controls are more highly connected in the hub nodes compared to the maltreated children.

\section{Shortest path length}
\index{path length}
\index{graph!path length}

\begin{definition}
In a weighted graph, the {\em shortest path} between two  nodes in a graph is a path whose sum of the edge weights is minimum. The {\em path length} between two nodes in a graph is the sum of edge weights in a shortest path connecting them \citep{bullmore.2009}. 
\end{definition}
When the weighted graph becomes binary, the path length is the number of edges in the shortest path. The shortest path is traditionally computed using Dijkstra's algorithm, which is a greedy algorithm with runtime $O(|V|^2)$ invented by E. W.  Dijkstra. The algorithm builds a {\em shortest path tree} from a root node, by building a set of nodes that have minimum distance from the root. The algorithm requires two two sets $\mathcal{S}$ and $\mathcal{N}$. $\mathcal{S}$ contains nodes included in the shortest path tree and $\mathcal{N}$ contains nodes not yet included in the shortest path tree. At every step of the algorithm, we find a node which is in $\mathcal{N}$ and has minimum distance from the root.

{\em Functional integration} in the brain is the ability to combine information from multiple brain regions \citep{lee.2018.TBME}. A measure of this integration is often based on the concept of path length. Path length measures the ability to integrate information flow and functional proximity between pairs of brain regions \citep{sporns.2004,rubinov.2010.NI}. When the path length becomes shorter, the potential for functional integration increases.

\section{Clustering coefficient} 
\index{graph!clustering coefficient}
\index{clustering coefficient}

Clustering coefficient describes the ability for functional segregation and efficiency of local information transfer.

The clustering coefficient of a node measures the propensity of pairs of nodes to be connected to each other if they are connected to another node in common \citep{newman.2001, watts.1998}. There are two different definitions of the clustering coefficients but we will use the one originally given in \citep{watts.1998}. 

\begin{definition} 
\label{def:cc}
Clustering coefficient $c_p$ at node $p$ is  fraction of the number of existing connections between the neighbors of the node divided by the number of all possible connections of the graph. 
\end{definition}

Let $k_p$ be the node degree at $p$. At most $k_p(k_p-1)/2$ edges can exists among $k_p$ neighbors if they are all connected to each other.  The {\em clustering coefficient} $c_p$ of the node $p$ is the fraction of allowable edges that actually exists over theoretical limit $k_p(k_p-1)/2$, i.e.,
$$c_p =  \frac{\mbox{ number of edges among neighbors of } p}{     k_p(k_p-1)/2}.$$
The {\em overall} clustering coefficient of a graph $G$, $c(G)$, is simply the average of the clustering coefficient $c_p$ over all nodes, i.e.,
\bq c(G) = \frac{1}{ |V|  } \sum_{p \in V} c_p, \label{eq:cG}\eq
where $|V|$ is the total number of nodes in the graph. Note that $0 \leq c(G) \leq 1$. 

\begin{figure}[t]
\centering
\includegraphics[width=1\linewidth]{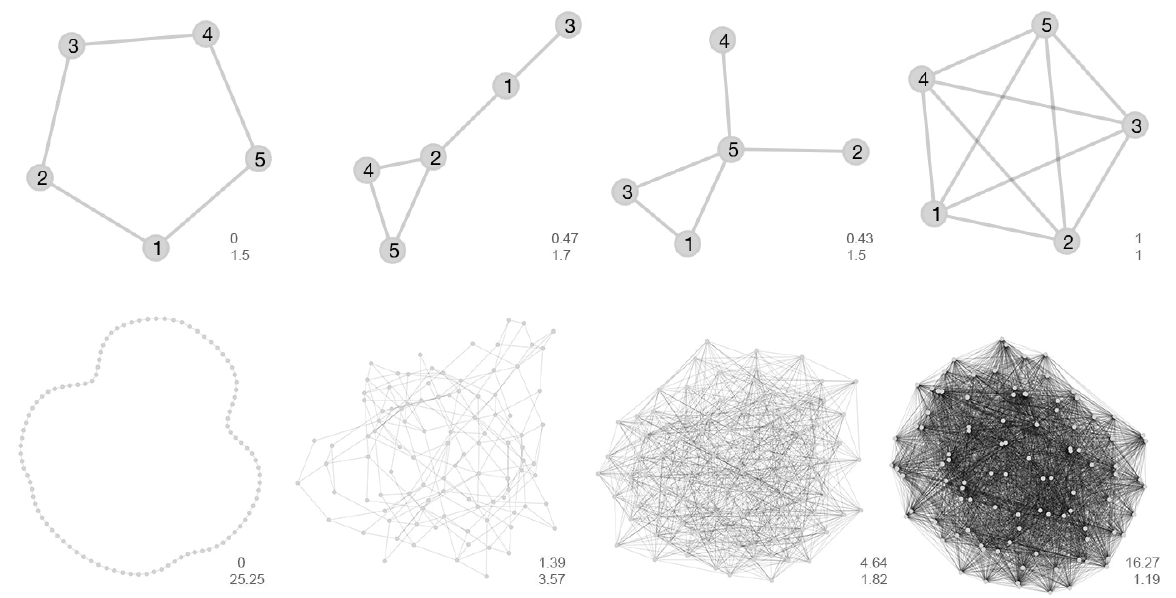}
 \caption{The numbers are the average clustering coefficients (top) and average path lengths (bottom). More complete the network becomes, it has shorter average path length and larger clustering coefficients.}\label{fig:graph-clustering}
\end{figure}

Random graphs are expected to have smaller clustering coefficient compared to more structured one \citep{sporns.2004}. For a complete graph, where all nodes are connected to each other, $c(G)$ obtains the maximum 1 and tends to zero for a random graph as the graph becomes large (Figure \ref{fig:graph-clustering}) \citep{newman.2006}. Definition \ref{def:cc} is biased for a graph with low degree nodes due to the factor $k_p(k_p-1)$ in the denominator. Unbiased definition of the clustering coefficient is computed by counting the total number of paired nodes and dividing it by the total number of such pairs that are also connected. Compared to random networks, the brain network is known to have higher clustering coefficient and shorter path length. These are the characterization of {\em small-world networks} \citep{sporns.2004}.

\section{Small-worldness} 
\index{graph!small-worldness}
\index{small-worldness}

If most nodes  can be connected in a very small number of paths, the network is said to be {\em small-world}. Small world networks have dense short range connections with relatively small number of long range connections \citep{watts.1998}. The small-worldness of networks is usually defined with clustering coefficient $c(G)$ and average path length $\mathbb{E} l$.

Let $l$ be the shortest path between two nodes,  and let $|V|$ be the number of nodes in a graph. If we take the average of $l$ for every pairs of nodes and over all realizations of the randomness in the model, we have the mean path length $\mathbb{E} l$. The mean path length $\mathbb{E} l$ measures the overall navigability of a network. $\mathbb{E} l$ is related to the diameter of the network, which is the maximum $l$ (longest path) \citep{newman.2003}.

 A network $G$ is considered to be a small-world network if it meets the following criteria:
\bq \gamma &=& \frac{c(G)}{c(R)} \gg 1\\
\lambda &=& \frac{\mathbb{E} l(G)}{\mathbb{E} l(R)} \approx 1\\
\sigma &=& \frac{\gamma}{\lambda} >1,
\eq
where $R$ denotes a random network that preserves the number of nodes, edges, and node degree distributions present in $G$. Often hundreds of random networks needed to be simulated for each network $G$ to obtain stable estimates for the average path lengths and clustering coefficient. Network small-worldness is often quantified by ratio $\sigma$.

For regular lattices, $\mathbb{E} l$ scales linearly with the number of nodes $|V|$ while for random graphs, $\mathbb{E} l$ is proportional to $\ln |V|$ \citep{newman.1999, watts.1998}. The small-world network is somewhere between regular lattice and random graphs. Therefore, the small-worldness is mathematically  expressed as \citep{song.2005}:
\bqn \mathbb{E} l \sim \ln |V|. \label{eq:connectivity-small}\eqn
The relation (\ref{eq:connectivity-small}) links over the size of the graph to the number of nodes and implies that as $|V|$ increases, the average path length is bounded by the logarithm of $|V|$. (\ref{eq:connectivity-small}) can be rewritten as
$$|V| \sim e^{\mathbb{E} l}.$$
(\ref{eq:connectivity-small}) implies that the small-world networks are not self-similar, since self-similarity requires a power-law relation between $l$ and $|V|$. However, \citep{song.2005} was able to show that the model (\ref{eq:connectivity-small}) might be biased for inhomogenous networks. Using a scale-invariant renormalization procedure through the box counting method, \citep{song.2005} was able to show diverse complex networks are in fact self-similar. 

\section{Fractal dimension}  
\index{fractal dimension}

The fractal dimension is often used to measure self-similarity. Benoit B. Mandelbrot named the term {\em fractal} in 1960 \citep{mandelbrot.1982}. Mathematically, a fractal is a set with non-integral Hausdorff dimension \citep{hutchinson.1981}. While classical geometry deals with objects with integer dimension, fractals have non-integral dimension. Fractals have infinite details at all points of the object, and have self-similarity between parts and overall features of the object. The smaller scale structure of fractals are similar to the larger scale structure. Hence, fractals do not have a single characteristic scale. Many anatomical objects, such as cortical surfaces and the cardiovasular system, are self-similar. The complexity of such objects can be quantified using the fractal dimension (FD). The main  question is if brain networks exhibit the characteristic of self-similarity. To answer this question, we need to compute the FD.

\begin{figure}[t]
\begin{center}
\includegraphics[width=1\linewidth]{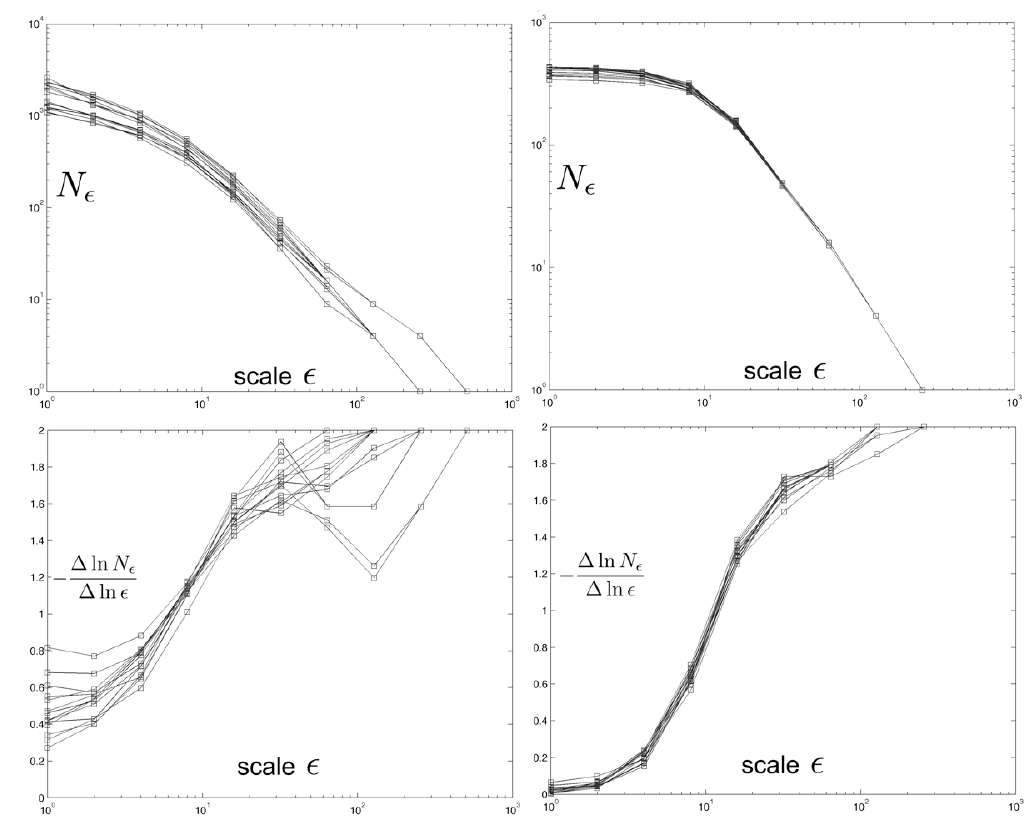}
 \caption{The plots of $\ln N_{\epsilon}$ and $-\frac{\Delta \ln N_{\epsilon}}{\Delta \ln \epsilon}$ over scale $\ln \epsilon$. The first column is  for 14 normal control subjects and the second column is for 14 Erd\"{o}s-R\'{e}nyi  random graphs $G(200,0.01)$. The FD characteristic of the random graph is different from the brain networks.} 
 \label{fig:connectivity-FDplot} 
 \end{center}
\end{figure}

Let $\epsilon$ be the scale and $N_{\epsilon}$ be the number of self-similar parts that can cover the whole structure. Then FD is defined as
$$ \mbox{ FD } = \lim_{\epsilon \to 0} \frac{\ln N_{\epsilon}}{\ln \frac{1}{\epsilon}}.$$
In practice, we cannot compute the limit as $\epsilon$ goes to zero for real anatomical objects so we resort to the box-counting method \citep{hutchinson.1981, mandelbrot.1982}. For $k$ different scales $\epsilon_1, \cdots, \epsilon_k$, we have corresponding number of covers $N_{\epsilon_1}, \cdots, N_{\epsilon_k}$. Then we draw the log-log plot of $(N_{\epsilon_i}, 1/\epsilon_i)$ and fit a linear line in a least squares fashion. The slope of the fitted line is the estimated FD. We can also estimate the FD locally using finite differences with neighboring measurements as
$$FD = -\frac{\Delta \ln N_{\epsilon}}{\Delta \ln \epsilon},$$
where $\Delta$ is the second order finite difference (Figure \ref{fig:connectivity-FDplot}).

For networks, we avoided using the box-counting method to the space where the network is embedded and the graph itself \citep{song.2005}. Rather we have applied the method to the structure that defines link connections, i.e., adjacency matrix. The use of the adjacency matrix simplifies a lot of computation. The Erd\"{o}s-R\'{e}nyi  random graph $G(n,p)$ is defined as a random graph generated with $n$ nodes where two nodes are connected with probability $p$ (Figure \ref{fig:connectivity-FDplot}) \citep{erdos.1961}. The FD characteristic of the brain networks is different from that of random graphs.  \\

{\em Other graph theory features.} Although we do not discuss them in detail here, other various popular graph theoretic measures are also proposed: entropy \citep{sporns.2000},  hub centrality \citep{freeman.1977} and modularity. A review on various graph measures can be found in \citep{bullmore.2009}. The centrality of a node measures the number of shortest paths between all other nodes that pass through the given node, and motivated in part by modeling social network \citep{freeman.1977}. Nodes with high centrality, which are likely to be with high-degree, are called hubs. The modularity of a network measures the number of components or modules and related to hierarchical clustering \citep{girvan.2002,lee.2012.TMI}.

\section*{Acknowelgement}
This study was in part supported by NIH grants R01 EB022856 and R01 EB028753 and NSF grant MDS-2010778. 
\bibliographystyle{plainnat}
\bibliography{reference.2021.02.25}

\end{document}